\documentclass{aa}  

\usepackage{tablefootnote}
\usepackage{graphicx}
\usepackage{multirow}
\usepackage{xspace}
\usepackage{xcolor}
\usepackage{siunitx}
\usepackage{afterpage}
\usepackage{array}
\usepackage{verbatim}
\usepackage{makecell}
\usepackage{soul}
\setbool{@fleqn}{false}
\usepackage{txfonts}
\usepackage[breaklinks,colorlinks,urlcolor=blue,citecolor=blue,linkcolor=blue]{hyperref}

\usepackage[flushleft]{threeparttable}

\makeatletter
\renewcommand*\aa@pageof{, page \thepage{} of \pageref*{LastPage}}
\makeatother

\def\mH{m_\mathrm{H}}
\def\chisqred{\ensuremath{\chi^2_\mathrm{red}}\xspace}

\newcommand{\orcidBildli}[1]{\raisebox{.3em}{\href{https://orcid.org/#1}{\includegraphics[width=0.7em]{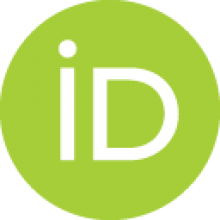}}}}

\begin{document} 

   \title{Resolving protoplanetary H$\alpha$ emission with the high-resolution spectrograph RISTRETTO}

   \author{
       J.~W.~Blackman
       \inst{\ref{Bern}}\orcidBildli{0000-0001-5860-1157}
  \and
       C.~Mordasini
       \inst{\ref{Bern}}\orcidBildli{0000-0002-1013-2811}
  \and
       G.-D.~Marleau
       \inst{\ref{Bern},\ref{UDE},\ref{MPIA}}\orcidBildli{0000-0002-2919-7500}
  \and
       C.~Lovis
       \inst{\ref{Genf}}\orcidBildli{0000-0001-7120-5837}
  \and
       M.~Bugatti
       \inst{\ref{Genf}}
  \and
       Y.~Aoyama
       \inst{\ref{SYSU}}\orcidBildli{0000-0003-0568-9225}
  \and
       N.~Blind\inst{\ref{Genf}}
      }

   \institute{Division of Space Research \&\ Planetary Sciences,
Physics Institute, University of Bern, Gessellschaftsstr.~6, 3012 Bern, Switzerland\label{Bern}
            \and
             Observatoire de Genève, University of Geneva, 51 chemin de Pegasi 1290 Versoix, Switzerland
             \label{Genf}
             \and
             Fakultät für Physik, Universität Duisburg--Essen, Lotharstraße 1, 47057 Duisburg, Germany
             \label{UDE}
             \and
             Max-Planck-Institut f\"ur Astronomie,
            K\"onigstuhl 17,
            69117 Heidelberg, Germany
            \label{MPIA}
            \and
             School of Physics and Astronomy, Sun Yat-sen University, Guangdong 519082, People’s Republic of China
             \label{SYSU}
             }

   \date{Received -- / Accepted --}

  \abstract
  {Thousands of exoplanets have been discovered to date, but only a handful of them are young planets in the later stages of their formation. H$\alpha$ emission is a strong tracer of this process, but it is difficult to isolate the planetary H$\alpha$ signal from the stellar signal due to the small angular separation. RISTRETTO is a high-resolution ($R\sim\textrm{140,000}$) integral-field unit spectrograph with an attached coronagraph and extreme adaptive system that is currently under construction. Used in tandem with the VLT, its resolving power and seven-spaxel layout is designed to characterise the atmospheres of exoplanets in reflected light. It can also be leveraged to obtain spatially and spectrally resolved observations of planetary H$\alpha$ emission lines. Using a combination of synthetic and real observations, we determined the detection limits for theoretical observations of protoplanets with on-sky separations of 37--600~mas (the equivalent of 4--70~AU at 113~pc). We used theoretical BT-Settl photospheric spectra, observed HARPS spectra of PDS70, and a model for the line emission at the planetary accretion shock to estimate the planetary H$\alpha$ signal. Models with the PyEchelle simulator show that the H$\alpha$ emission from theoretical objects similar to the four known protoplanet candidates (PDS70b, PDS70c, WISPIT2b, and 2MJ1612b) can be spectrally resolved with VLT/RISTRETTO down to a line flux between $F\sim(1.4$--$2.7)\times10^{-16}~\mathrm{erg\; s^{-1}\; cm^{-2}}$ with a 1~hour exposure. We show that the planetary H$\alpha$ line can be spatially resolved from that of its stellar host and that the shape of the line can be used to constrain the pre-shock density ($n_0$) and velocity ($v_0$), thereby enabling new constraints on the accretion rate and flow geometry. This study and the instrument act as a pathfinder for the next generation of high-resolution spectrographs such as ELT/PCS and ELT/ANDES.}

   \titlerunning{Resolving protoplanetary H$\alpha$ with RISTRETTO}
   \authorrunning{J.W.\ Blackman et al.}
   \maketitle

\section{Introduction}
Planets are believed to form within conglomerations of gas known as protoplanetary discs \citep{Hayashi1981}. One limitation in our ability to understand gas giant formation at present is the lack of direct imaging observations of young ($< 10$~Myr) protoplanets embedded in these discs that are still undergoing formation \citep{Hernandez2008}, despite several efforts to observe them \citep{Cugno2019,Follette2023}. A number of candidate systems have been found, most notably AB Aurigae b \citep{Currie2022b}. However, until recently, the only unambiguous detection was PDS 70 \citep{Keppler2018, Haffert2019}, which consists of at least two embedded protoplanets, b and c, and which has been observed over a variety of wavelengths \citep{Close2025a}. Two other strong candidates are the accreting super Jupiters WISPIT 2b \citep{Close2025} and 2MJ1612b \citep{Li2025a}, both with ages of $\sim$5 Myr. The detection of these young objects is a way to place constraints on accretion rate history (i.e. how and when solids and gas are accumulated together to form planets) and thereby constrain formation models. The most accessible way to achieve this is by measuring the H$\alpha$ line profiles of these objects. 

H$\alpha$ (656.28~nm in air) emission occurs when hydrogen infalls on the protoplanet and/or its surrounding circumplanetary disc and undergoes a shock \citep{Aoyama2018}. A portion of the H$\alpha$ emission might come from accretion funnels or possibly from magnetic processes in the chromosphere of the planet \citep{Manara2017,Eriksson2020}.
According to the models of \cite{Aoyama2018}, the shape of the curve can be described by the gaseous hydrogen number density, $n_0$, and gas infall velocity, $v_0$. The spectral widths of the emission line, estimated at various fluxes (e.g. 10\%, 50\%) can be used to estimate the accretion rate.
By observing the H$\alpha$ line at (or beyond) a resolution at which the widths can be accurately determined, we  can constrain the physical processes responsible for the emission. \cite{Aoyama2020} predicts this line width to be fully resolved at spectral resolution of $R\equiv\lambda/\Delta\lambda\sim\textrm{18,000}$. Observing at an even higher resolution ($R > \textrm{100,000}$) would enable the detection of individual spectral features, possibly coming from photospheric emission or from absorption due to extinction from the gas in the surrounding circumplanetary and circumstellar discs \citep{Marleau2022}.

\begin{figure*}[t]
\sidecaption
\includegraphics[width=0.70 
 \textwidth]{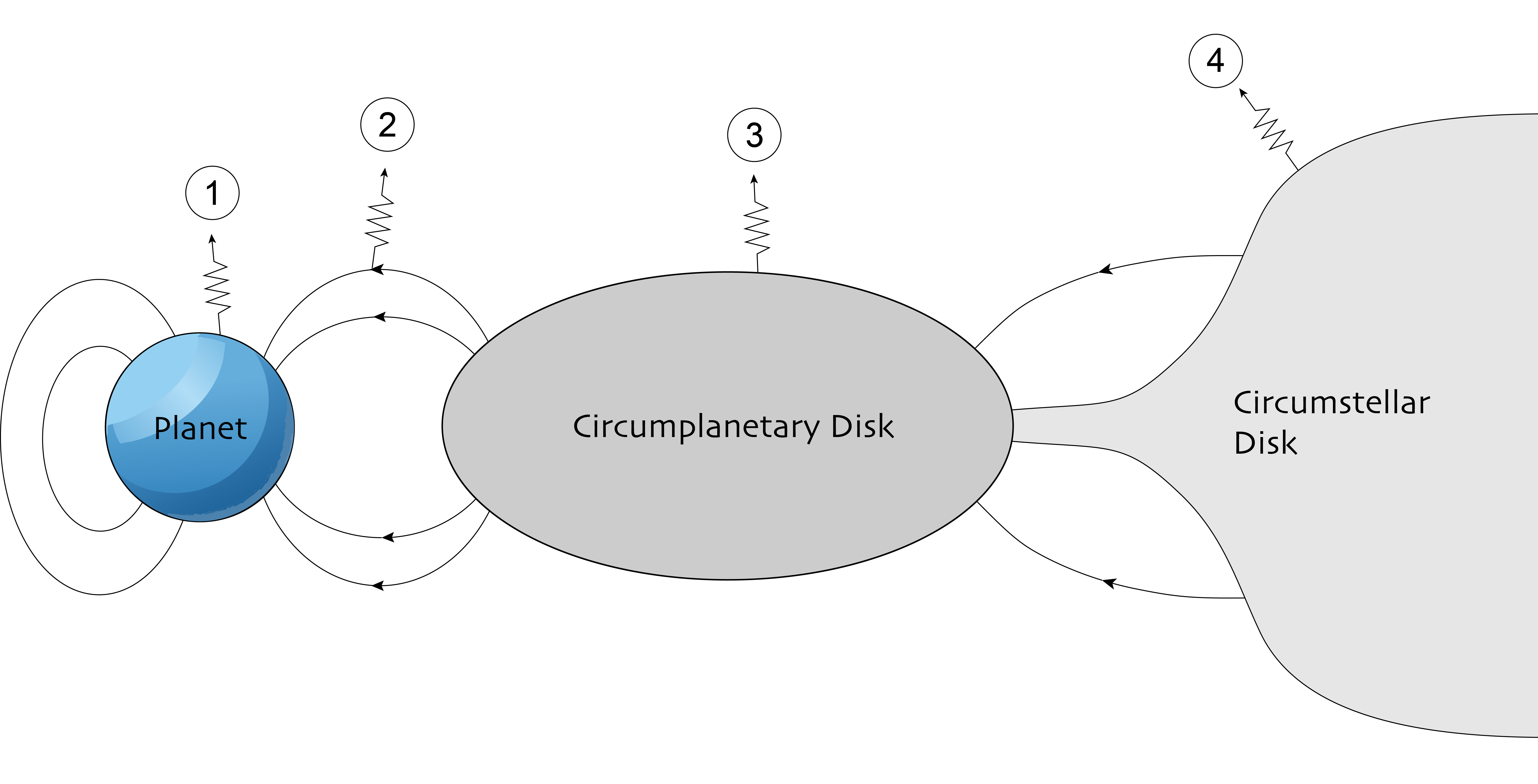}
\caption{Schematic showing possible sources of H$\alpha$ emission in the region surrounding forming protoplanets. These include: (1) shock from infalling gas at the planet surface; (2) accretion funnels; (3) gas incident on the circumplanetary disc; and (4) from reflected stellar light.
\label{fig:emission_schematic}}
\end{figure*}

In this paper, we discuss the potential of RISTRETTO\footnote{\url{https://ristretto.astro.unige.ch}} (high-Resolution
Integral-field Spectrograph for the Tomography of Resolved
Exoplanets Through Timely Observations) as a tool for spatially and spectrally resolving this H$\alpha$ emission from young protoplanets \citep{Lovis2024}. RISTRETTO is a pathfinder instrument for ELT/ANDES \citep{Marconi2024} and ELT/PCS \citep{Kasper2021} and is comprised of a combination of a visible-wavelength spectrograph, coronagraph, and an extreme adaptive-optics (XAO) system. It is designed to enable ground-based characterisation of the atmosphere of Proxima~b and other rocky terrestrial exoplanets in reflected light for the first time. With a planned resolving power of $R\sim\textrm{140,000}$, visible-light sensitivity at 620--840~nm and a seven-spaxel, single-mode fibre design, it will be capable of isolating the planetary H$\alpha$ line of PDS70 from its stellar host. Since each spaxel covers 37~mas on sky, we expect this to be achievable for protoplanets in the closest star-forming regions ($\sim$140~pc).

In this paper, we focus on the four protoplanet cases with solid H$\alpha$ detections (PDS70b, PDS70c, WISPIT2b and 2MJ1612b) and use them to estimate RISTRETTO detection limits. We restrict our investigation to these specific planetary-mass objects located in gaps in their natal protoplanetary disc. In Sect.~\ref{sec:currentstate}, we discuss the current state of H$\alpha$ observations with respect to theoretical predictions. In Sect.~\ref{sec:hellomynameisRISTRETTO}, we discuss the instrument specifications and in Sect.~\ref{sec:Hasens}, we demonstrate the sensitivity and observational potential of the instrument using a combination of real and theoretical input spectra. Our conclusion and discussion are presented in Sect.~\ref{sec:concdisc}.

\section[State of Play: Accreting Protoplanets in H alpha]{State of play: Accreting protoplanets in H$\alpha$}
\label{sec:currentstate}

\subsection{Theoretical rationale}

H$\alpha$ emission is the most commonly used accretion tracer due to its brightness, but our understanding of the mechanisms and sources of the accretion is incomplete. One explanation is that this emission is produced in the hot ($\sim10^4$~K) non-equilibrium post-shock region, with gas infalling onto the protoplanet and surrounding circumplanetary disc (see items~1 and~3 in our Fig.~\ref{fig:emission_schematic} and the schematic Fig. 2 of \citealt{Marleau2022}). This gas is thought to originate in the circumplanetary disc (CPD) and flows nearly vertically before supersonically colliding with the surface of the disc, resulting in a strong shock. This shock emission is a viable mechanism that has been shown to contribute to the emission seen for a number of objects \citep{Betti2022, Demars2023, Viswanath2024}.

Accretion funnels \citep{Thanathibodee2019} 
might contribute to the total H$\alpha$ flux in the planetary case. If they are indeed present, this would result in the observed H$\alpha$ line deviating from the expected Gaussian profile. \cite{Aoyama2021a} noted that asymmetries across the centre of the line and broadening at the shorter wavelengths could indicate the presence of funnel emission. Although it might be asymmetric, the flux still changes monotonically. As such, the most complex feature that could be seen is a flattening at the peak $-$ and not, for example, up-down features such as a double peak. Two other possible sources are chromospheric emission from the magnetic field surrounding the protoplanet, for which there are few predictions of line shapes in the literature \citep{Demars2026}, and contamination from reflected H$\alpha$ emission from the host star (e.g. \citealp{Haffert2019,Aoyama2021a,Zhou2023}).

\begin{figure*}[htbp]
  \centering
  \includegraphics[width=0.46\textwidth]{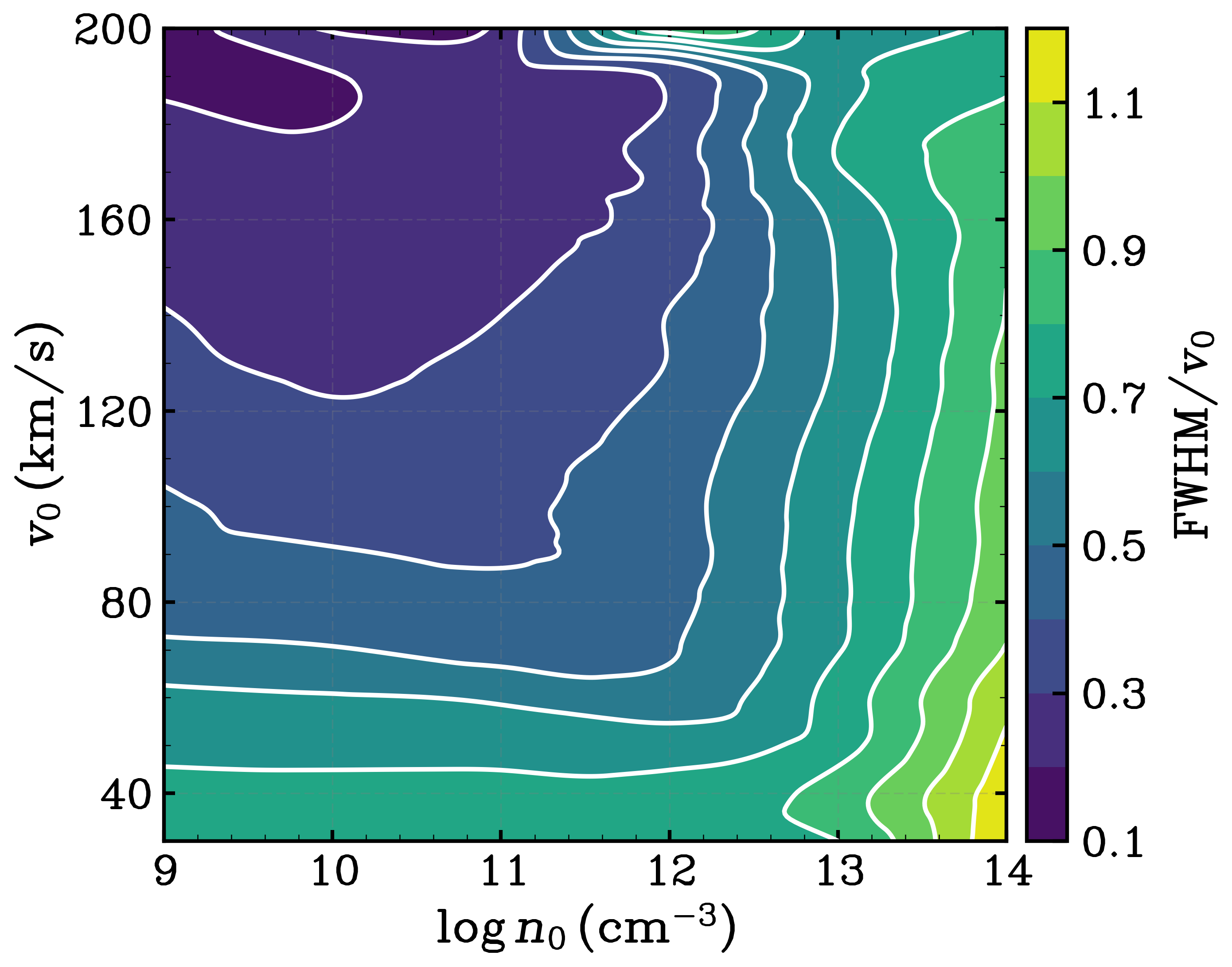}
  ~~~~
  \includegraphics[width=0.46\textwidth]{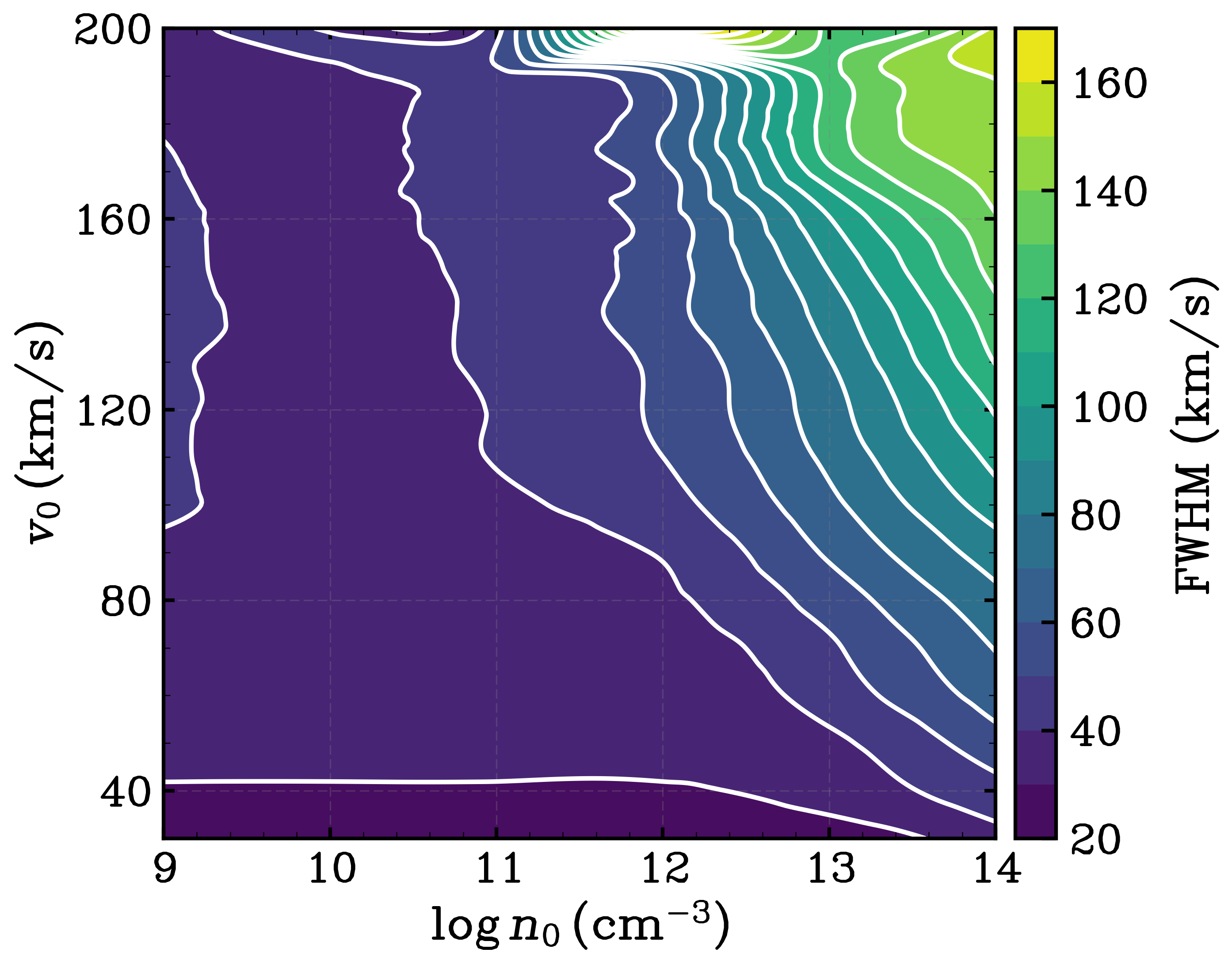}
  \caption{Dependence of the relative ($\textrm{FWHM}/v_0$, left) and absolute FWHM (right) H$\alpha$ line width on the pre-shock hydrogen number density, $n_0$, and velocity, $v_0$, in the shock-emission model of \cite{Aoyama2018}.}
  \label{fig:halpha_v0n0}
\end{figure*}

Until recently, any inference using the empirical relation between the 10$\%$ full width of the H$\alpha$ line and accretion rate was based on stellar models of T Tauri stars \citep{Natta2004}. The stellar case is distinct from the planetary scenario because the higher host masses lead to higher temperatures ($T > 10^5$~K) that inhibit line emission \citep{Aoyama2021a}. The overall limited understanding of the planetary case motivated the construction of the \cite{Aoyama2018} model, which is a 1D radiation-hydrodynamic code describing the hydrogen-line emission produced below the accretion shock at the surface of the circumplanetary disc or the protoplanet (in the narrow disequilibrium cooling zone; see schematic in Fig.~2 of \citealt{Marleau2022}).

The H$\alpha$ luminosity can be described using pre-shock velocity ($v_0$), number density of hydrogen ($n_0$), and line-emitting flux, $F_{\mathrm{H}\alpha}$ \citep{Aoyama2018}. It is distributed over a planetary sphere radius, $R_P$, and includes corrections for extinction, $A_{\mathrm{H} \alpha}$, and the fractional area (i.e. the filling factor) responsible for the accretion flow, $f_{\mathrm{f}}$, via
\begin{equation}
L_{\mathrm{H} \alpha}=4 \pi R_{\mathrm{P}}^2 f_{\mathrm{f}} F_{\mathrm{H} \alpha}(v_0,n_0) 10^{-0.44A_{\mathrm{H} \alpha}}.
\label{eq:lum1}
\end{equation}
Following \citet{Aoyama2019} and \cite{Hashimoto2020}, this can be rewritten as
\begin{equation}
L_{\mathrm{H} \alpha}= \frac{G M_{\mathrm{P}} \dot{M}}{R_{\mathrm{P}}} \frac{F_{\mathrm{H} \alpha}\left(v_0, n_0\right)}{\frac{1}{2}\mu^{\prime} n_0 v_0^3} 10^{-0.4A_{\mathrm{H}\alpha}},
\label{eq:lum2}
\end{equation}
using the relationships, 
\begin{align}
\dot{M} &=4 \pi R_{\mathrm{P}}^2 f_{\mathrm{f}} \mu^{\prime} n_0 v_0\label{eq:accretionrate},\\
M_{\mathrm{P}} &=R_{\mathrm{P}}v_{0}^{2}/2G,
\end{align}
where $M_P$ is the planetary mass, $G$ is the gravitational constant, and $\mu^{\prime}=\mH/X$ is the mean weight per hydrogen nucleus, with $\mH$ as the mass of a hydrogen nucleus and $X=0.738$ the solar-composition mass fraction of hydrogen (see Eq.~A6 of \citealt{Aoyama2020}). Equation \ref{eq:lum2} is written as three components representing (from left to right) the accretion luminosity, line emission efficiency (line flux per accretion energy per unit area per unit time), and an extinction correction.

Equation~\ref{eq:accretionrate} indicates that by fitting an observationally obtained H$\alpha$ line shape, we can determine the accretion rate from the line width; however, this is also dependent on the 10\% width and estimates of $R_P$, $f_f$, and $A_{H\alpha}$. Equations~\ref{eq:lum1} and~\ref{eq:lum2} show this relationship between the line flux and ($v_0,n_0$), while Fig.~\ref{fig:halpha_v0n0} gives the full width at half maximum (FWHM) as a function of ($v_0,n_0$). This figure shows a degeneracy that could be broken by comparing different percentage line widths (e.g. 10\%, 50\%, 90\%), as shown in \citet{Aoyama2019}; however, this is not guaranteed in all cases. 

\begin{figure}[t]
\centering
\includegraphics[width=0.49\textwidth]{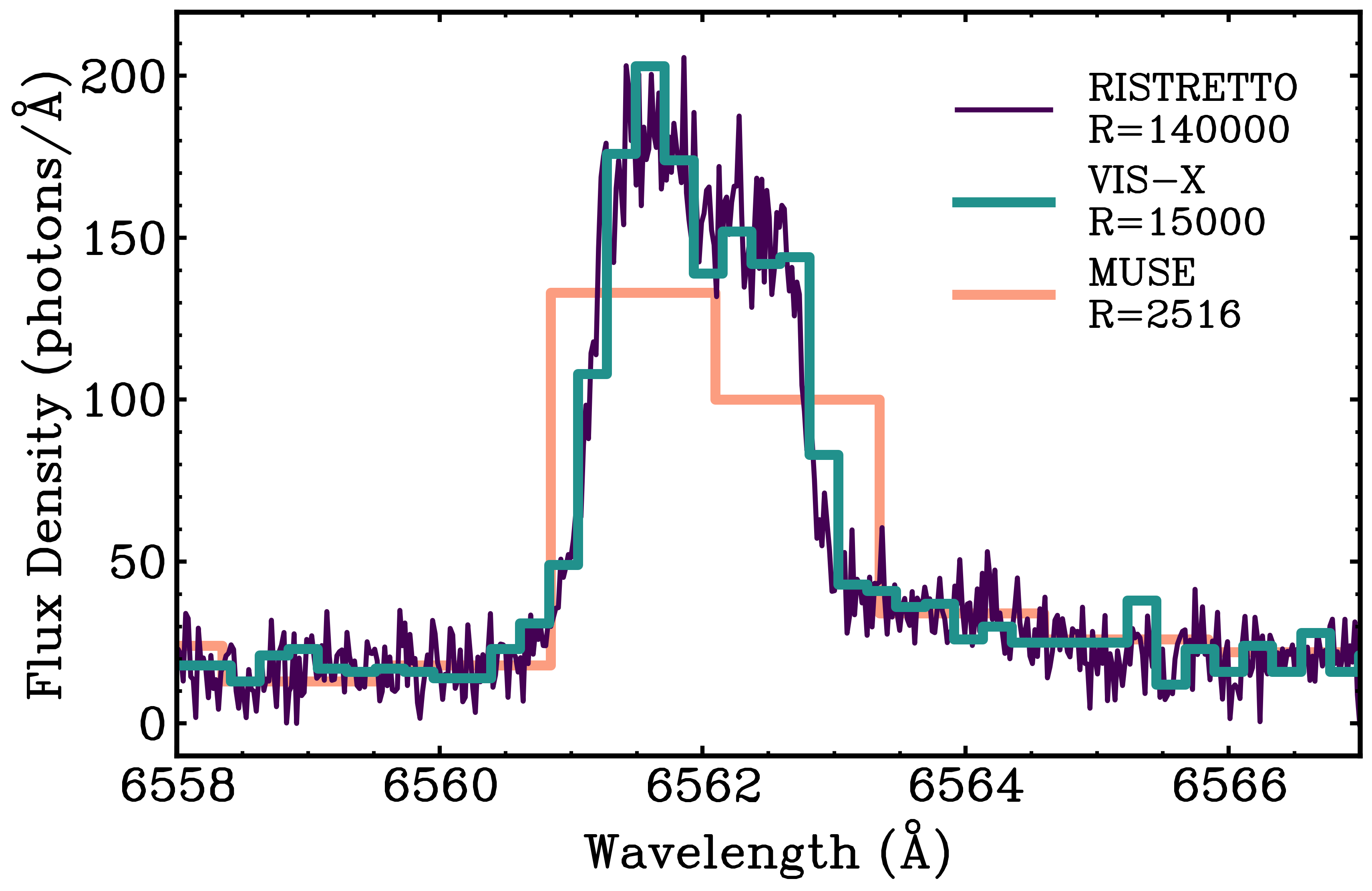}
\caption{Comparison of the resolving power of RISTRETTO, VIS-X \citep{Haffert2021}, and MUSE \citep{Xie2020} derived from a high-resolution a synthetic H$\alpha$ emission profile using the model from \cite{Aoyama2018} including simulated noise added with PyEchelle \citep{Sturmer2018}. The simulated 1-hour observation has the parameters of 2MJ1612b. This spectrum was resampled to the resolution of VIS-X and MUSE at the Nyquist rate (2 pixels per resolution element).
}
\label{fig:discovery}
\end{figure}

One advantage of RISTRETTO is its high spectral resolution. Figure~\ref{fig:discovery} compares its resolving power to that of MUSE and VIS-X. RISTRETTO allows both the star and planet to be spatially separated into different spaxels, while the line width and absorption features can be fully characterised. For simplicity, in this work, we ignore the possible extinction, $A_{\mathrm{H\alpha}}$, from the surrounding disc. Over a narrow wavelength range and for the moderate accretion rates of our targets, the extinction is either negligible or only reduces the flux but it does not introduce spectral features, depending on the exact properties of the dust grains. For a detailed treatment of extinction for different accretion flows, we refer to \cite{Marleau2022}.

\subsection{Observations}

Despite the existence of prior campaigns aimed at detecting other sources \citep{Cugno2019, Zurlo2020, Xie2020, Flores-Rivera2023,Follette2023}, observational evidence of protoplanets (planetary-mass objects in the disc around their host star) currently undergoing accretion is limited to PDS 70 b and c \citep{Keppler2018, Muller2018, Wagner2018, Haffert2019, Zhou2021}, WISPIT2b \citep{Close2025}, 2MJ1612b \citep{Li2025a}, and potentially AB Aur b \citep{Zhou2022,Currie2025, Shibaike2025}. However, there is still some discussion around the possibility that some of the emission from the latter could be due to scattered light from the disc \citep{Zhou2023}. Delorme~1(AB)b \citep{Delorme2013,Eriksson2020, M-acirc-lin2025, Demars2025} has had its emission detected in H$\alpha$ and H$\beta$ and has been confirmed as a young planet, but its large projected separation (1\farcs77) places it beyond the scope of this study.\footnote{LkCa15 has also been discussed, but it has been clearly shown that the H$\alpha$ signal is not from an accreting protoplanet \citep{Kraus2012, Sallum2015, Currie2019, Sallum2023a, Gardner2025, Swastik2026}.} This comparative dearth of confirmed detections is likely due to the faintness of targets \citep{Marleau2026a} in the nearby star-forming regions ($\sim$140~pc), the complexity of isolating the planetary signal from that of the star (see also \citealt{Ragusa2025}), selection effects \citep{Close2020}, and 
possible extinction of the planetary emission within the circumstellar disc (\cite{Cugno2025}; see also \cite{VanBoekel2017}. However, we note that these cited studies were focused on lower mass planets that would not be expected to generate emission lines). In this study, we restrict our investigation to the first four targets accessible at the latitude of the VLT ($26.62\degr$~S).

PDS70 hosts two confirmed planets, PDS70b and PDS70c (as well as a possible third, PDS70d), at separations of 20.6~and 34.5~AU, respectively (\citealp{Haffert2019}; see literature overview of system properties in \citealt{Shibaike2024}), corresponding to $176.8\pm25$ and $235.5\pm25$~mas on sky. Its host is a young, 5~Myr old K7 T Tauri star \citep{Hashimoto2015} surrounded by a gaseous, pre-transitional disc \citep{Espaillat2007}. The H$\alpha$ line flux of PDS70b is estimated to be in the range of $F = (2.28$--$10.4)\times10^{-16}\,\mathrm{erg\,s^{-1}\,cm^{-2}}$, while PDS70c is characterised by $F = (1.9$--$4.78)\times10^{-16}\,\mathrm{erg\,s^{-1}\,cm^{-2}}$ \citep{Haffert2019, Hashimoto2020, Close2025a}. For PDS70b, the flux variation of $8.1\times10^{-16}$ between 2022 and 2024 could indicate some variability in the underlying accretion processes \citep{Close2025a}. Table~\ref{table:halpha_flux} shows the current line flux estimates for both objects, along with values from \cite{Xie2020} based on the same data as \cite{Haffert2019}, but accounting for flux loss due to MUSE’s low (<20\%) Strehl. \cite{Hashimoto2020} is based on a re-reduction and re-analysis of the same data. While PDS 70 bc have have been observed extensively across the electromagnetic spectrum $-$ with, for instance, SPHERE/IFS \citep{Muller2018}, MUSE \citep{Haffert2019}, Keck/NIRC2 \citep{Wang2020}, and HST/WFC3/UVIS \citep{Zhou2021} $-$ the H$\alpha$ signal has not yet been observed in a study combining high resolution and a significant spatial separation from the host star.
\afterpage{
\begin{table*}
\vspace{40pt}
\centering
\begin{threeparttable}
\caption{Estimated H$\alpha$ line fluxes for observed protoplanets.}
\setlength{\tabcolsep}{6pt} 
\renewcommand{\arraystretch}{1.3}

\begin{tabular}{ccccc}
\hline
\noalign{\vskip 1pt}
\hline
Object & \makecell[c]{H$\alpha$ Line Flux \\[0.6mm] {\small [$10^{-16}$~erg cm$^{-2}$ s$^{-1}$]}} & Date & Instrument & Reference \\
\hline
\multirow{5}{*}{PDS70b} 
& $3.9\pm0.37$ & 20 June 2018 & VLT/MUSE & \cite{Haffert2019}\\
& $8.1\pm0.3$ & 20 June 2018 & VLT/MUSE & \cite{Hashimoto2020}\\
& $21.8 \pm 2.1$\textsuperscript{\textdagger} &20 June 2018 & VLT/MUSE & \cite{Xie2020}\\
& $10.4\pm1.6$ & 24 April 2022 & Mag AO-X & \cite{Close2025a}\\
& $2.28\pm0.26$ & 8 March 2023 & Mag AO-X & \cite{Close2025a}\\
& $3.64\pm0.87$ & 25 March 2024 & Mag AO-X & \cite{Close2025a}\\
\hline
\multirow{5}{*}{PDS70c} 
& $1.9\pm0.32$ & 20 June 2018 & VLT/MUSE & \cite{Haffert2019}\\
& $3.1\pm0.3$ & 20 June 2018 & VLT/MUSE & \cite{Hashimoto2020}\\
& $10.6\pm1.8$\textsuperscript{\textdagger} &20 June 2018 & VLT/MUSE & \cite{Xie2020}\\
& $3.3\pm1.5$ & 24 April 2022 & Mag AO-X & \cite{Close2025a}\\
& $2.04\pm0.21$ & 8 March 2023 & Mag AO-X & \cite{Close2025a}\\
& $4.78\pm0.46$ & 25 March 2024 & Mag AO-X & \cite{Close2025a}\\
\hline
\multirow{2}{*}{WISPIT~2b} 
& $13.8\pm3.3$ &  13 April 2025 & MagAO-X  & \cite{Close2025} \\
& $12.9\pm2.8$ &  16 April 2025 & MagAO-X  & \cite{Close2025} \\
\hline
\multirow{2}{*}{2MJ1612b} 
& $29.7\pm7.5$ & 13 April 2025 & MagAO-X & \cite{Li2025a} \\
& $8.2\pm3.4$ & 16 April 2025 & MagAO-X & \cite{Li2025a} \\
\hline
\end{tabular}

\begin{tablenotes}
\small
\item[]
\textsuperscript{\textdagger}Utilises the line fluxes of \cite{Haffert2019} but applies a factor of $\sim$5.5 to correct for flux losses due to the low Strehl ratio (<20\%).
\end{tablenotes}

\label{table:halpha_flux}
\end{threeparttable}
\end{table*}
\begin{table*}
\vspace{25pt}
\centering
\begin{threeparttable}
\caption{Physical properties of the four known accreting protoplanets accessible using VLT/RISTRETTO.}
\setlength{\tabcolsep}{6pt} 
\renewcommand{\arraystretch}{1.3}
\begin{tabular}{c
    >{\centering\arraybackslash}m{2.4cm}
    >{\centering\arraybackslash}m{2.4cm}
    >{\centering\arraybackslash}m{2.4cm}
    >{\centering\arraybackslash}m{2.4cm}}
\hline
\noalign{\vskip 1pt}
\hline
 & PDS70b\textsuperscript{1} & PDS70c\textsuperscript{2} & WISPIT 2b\textsuperscript{3} & 2MJ1612b\textsuperscript{4} \\
\hline
Mass [$M_\mathrm{Jup}$] 
& 2--4 & 1--3 & $5.3\pm1.0$ & 4 \\
Radius [$R_\mathrm{Jup}$] 
& $2.0\pm0.2$ & $1.60_{-0.19}^{+0.30}$ & 1.6 & 1.5 \\
Planet $T_\mathrm{eff}$ [K] 
& 1400 & 1300 & 1400 & 1200 \\
Planet $\log g$ 
& 4.0 & 4.0 & 3.7 & 3.7 \\
\hline
Separation [au] 
& 20.6 & 34.5 & 57.5 & $23.5\pm0.3$ \\
Separation [mas] 
& $176.8\pm25$ & $235.5\pm25$ & $309.5\pm1.5$ & $142\pm2$ \\
\hline
Age [Myr] 
& $5.4 \pm 1.0$ & $5.4 \pm 1.0$ & $5.1_{-1.3}^{+2.4}$ & 5--10 \\
Stellar mass [$M_\odot$]
& \textsuperscript{a}0.76 & \textsuperscript{a}0.76 & 1.1 & 0.7 \\
Stellar radius [$R_\odot$]
& \textsuperscript{a}1.26 & \textsuperscript{a}1.26 & 1.4 & 1.2 \\
Stellar $T_\mathrm{eff}$ [K]
& 4200 & 4200 & \textsuperscript{b}4400 & 3900 \\
Stellar $\log g$
& 4.8 & 4.8 & \textsuperscript{b}4.1 & 4.2 \\
\hline
\end{tabular}

\label{table:planetproperties}

\begin{tablenotes}
\small
\item[]
\textsuperscript{1}\cite{Wagner2018}, \cite{Haffert2019}, \cite{Hashimoto2020}, \cite{Zhou2021}, \cite{Wang2021};
\textsuperscript{2}\cite{Haffert2019}, \cite{Hashimoto2020}, \cite{Wang2021};
\textsuperscript{3}\cite{Close2025}, \cite{Capelleveen2025}; 
\textsuperscript{4}\cite{Li2025a};  
\textsuperscript{a} \cite{Muller2018}, \cite{Cridland2023} \textsuperscript{b} \cite{Baraffe2015} 
\end{tablenotes}

\end{threeparttable}
\end{table*}
\clearpage
}

In August 2025, the accreting protoplanet WISPIT2b was found around the star TYC 5709-354-1 \citep{Capelleveen2025}. In their study of this object, \cite{Close2025} reported a mass of $5.3\pm1.0\;M_\mathrm{{Jup}}$ and a radius of $1.6\;R_\mathrm{{Jup}}$. Detected with an H$\alpha$ line flux of  $F=(1.38\pm0.33)\times10^{-15}\,\mathrm{erg\,s^{-1}\,cm^{-2}}$, WISPIT2b sits at a projected separation of $r=309.48\pm1.56$~mas, which corresponds to 57.5~AU.  Observations a few days later measured $F = (1.29\pm0.28)\times10^{-15}\,\mathrm{erg\,s^{-1}\,cm^{-2}}$, which is statistically in agreement.

The fourth accreting planet was found in 2MJ1612 (2MASS J16120668–3010270), which has an inclined disc. In the large dust-depleted gap of 53~AU, observations with MagAO-X revealed an H$\alpha$ excess 142~mas from the star. This was interpreted as a 4$\;M_\mathrm{{Jup}}$ planet with a line flux, measured three days apart, of either $F = (30\pm8)$~or $(8\pm3)\times10^{-16}\,\mathrm{erg\,s^{-1}\,cm^{-2}}$ \citep{Li2025a}. The reason for this difference in flux is unclear. The physical properties of these and the other planets we use in our simulations are shown in Table~\ref{table:planetproperties}, while the range of H$\alpha$ fluxes are shown in Table~\ref{table:halpha_flux}. They are all super-Jupiters of similar radius (1.5--2$\;R_\mathrm{{Jup}}$) at relatively wide separations (23--58~AU).

\section{The RISTRETTO Instrument}
\label{sec:hellomynameisRISTRETTO}

RISTRETTO was born of the notion that coupling the SPHERE imager and ESPRESSO spectrograph (i.e. the two instruments of the Very Large Telescope, VLT) can aid in the effort to characterise the atmosphere of the temperate rocky exoplanet Proxima~b \citep{Lovis2016}. By virtue of being the closest exoplanet to Earth at a distance of 4.24~light years, Proxima~b is among the easiest to detect, despite the very challenging planet-star contrast ratio ($\sim10^{-7}$). This contrast ratio and the maximum planet--star separation of 37~mas informs a design that is explicitly intended to characterise the atmospheres of nearby habitable-zone exoplanets in reflected light \citep{Lovis2024}.

\begin{figure}[t]
\centering
\includegraphics[width=0.49\textwidth]{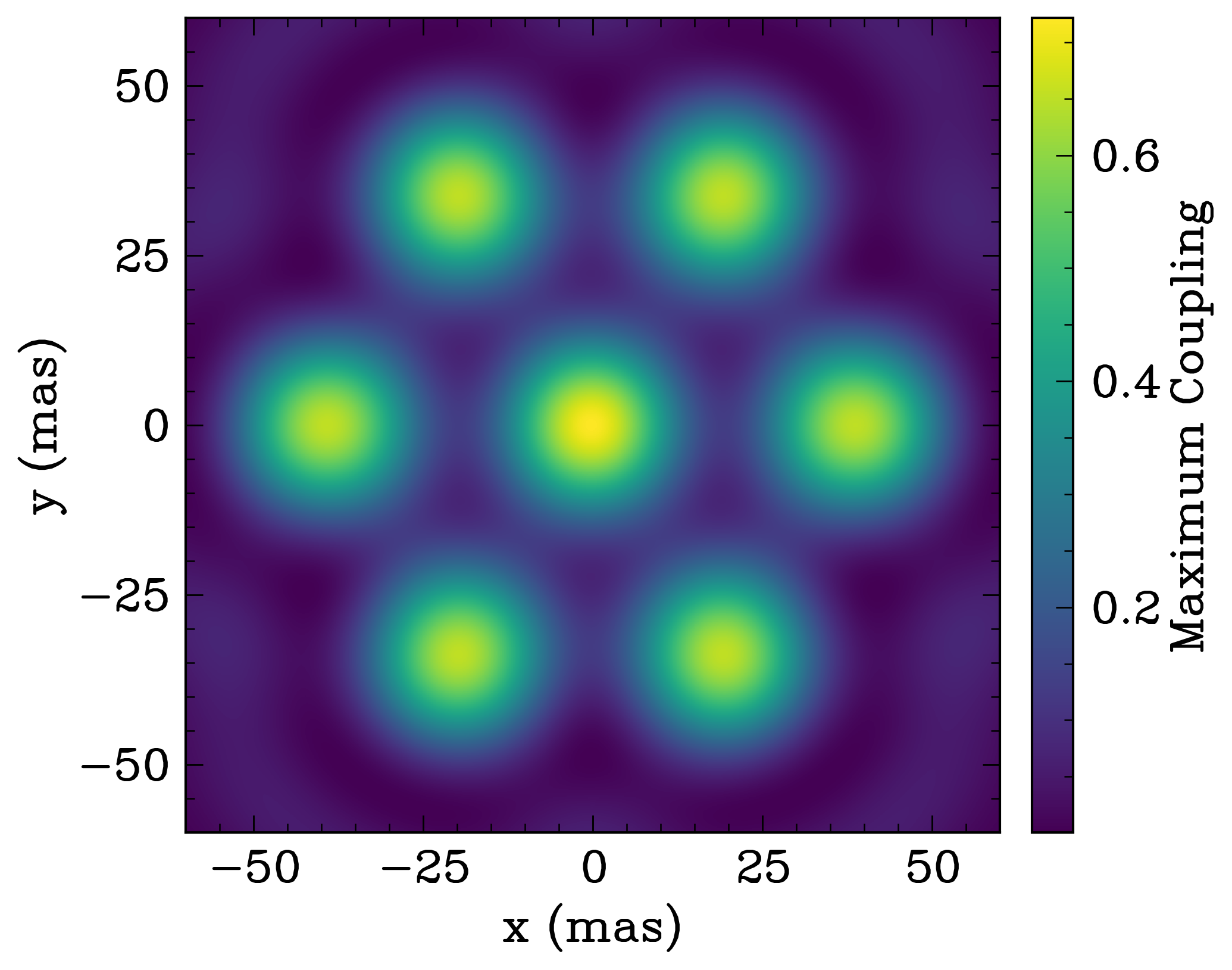}
\caption{Seven-spaxel layout of the RISTRETTO spectrograph. The colour gradient shows the sensitivity at different position offsets across the seven single mode fibres. The falloff between the spaxels is such that it is important to align science targets centrally in each spaxel.
\label{fig:ristretto}}
\end{figure}

\begin{figure}[t]
  \centering
  \includegraphics[width=0.48\textwidth]{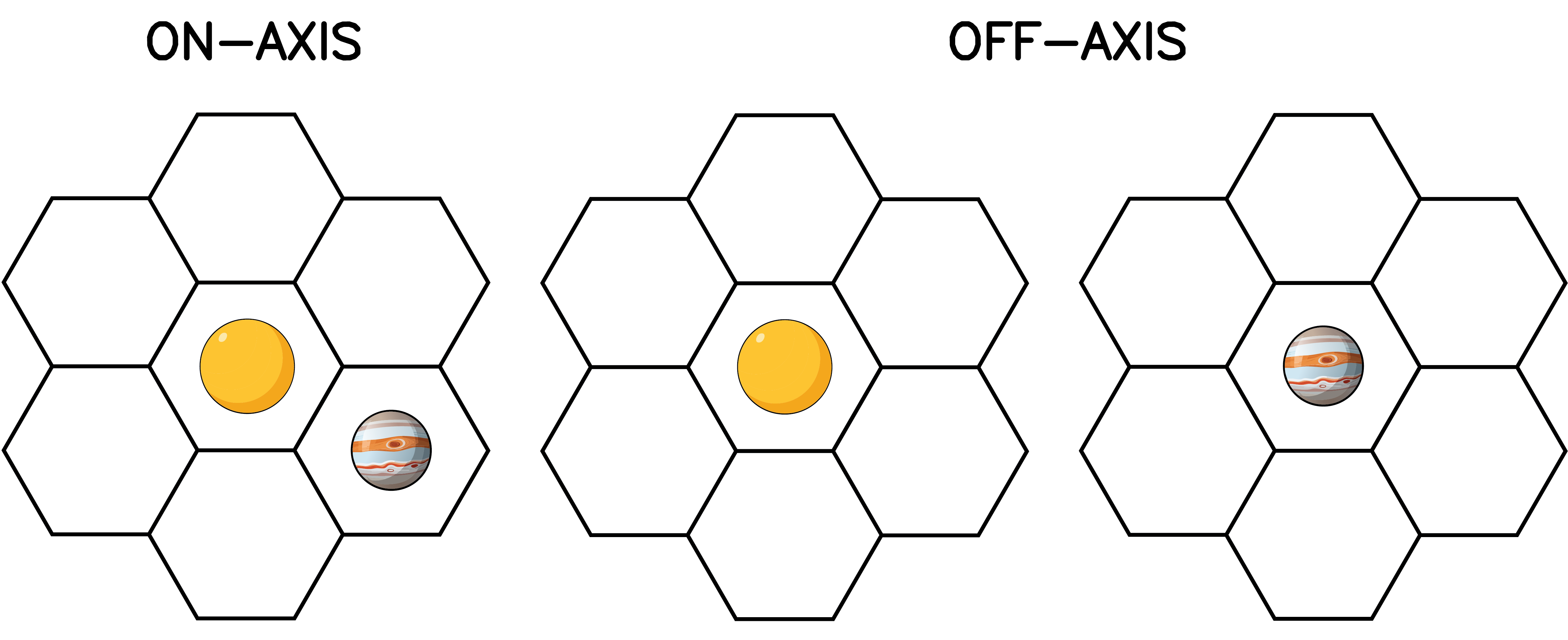}
  \caption{Comparison between the planned on-axis (left) and off-axis (centre and right) observations with RISTRETTO. The on-axis observations involve a single exposure of the star in the central spaxel and the planet in the external spaxel (37~mas separation). The off-axis observations involve sequential exposures of the star and then the planet in the central spaxel (separations of 100--600~mas).}
  \label{fig:ristretto_onffaxis}
\end{figure}

Planned as a visitor instrument on the VLT, RISTRETTO will consist of a front end containing an extreme AO system, a coronagraph-and-apodiser, and a back-end with a seven-spaxel integral-field unit (IFU). A spaxel (spatial pixel) is the smallest spatial component input into the spectrograph and, in this case, there are seven hexagonal lenslets with an on-sky size of 37~mas. These lead into single-mode fibres. Figure~\ref{fig:ristretto} presents the layout of these spaxels and the falloff in coupling (i.e. how much light is transmitted into each fibre) calculated from a simulation of the system throughput \citep{Blind2025}. The spaxel layout is such that for the default `on-axis' observations, the star is centred in the middle spaxel and the planet is centred on the external spaxel. Figure~\ref{fig:ristretto_onffaxis} gives a schematic example of both so-called on-axis and off-axis observations. The coronagraph suppresses the stellar halo in the external spaxel to an order of $1\times10^{-4}$ (see the discussion in Sect. \ref{subsection_onaxis}), thus enabling the planetary signal to be isolated.

As it is designed to operate at the diffraction limit of the telescope, \cite{Bugatti2025} predicted that the instrument could detect Proxima~b with 55 hours of VLT time and $\mathrm{O_2}$ and $\mathrm{H_2O}$ absorption lines after 85 hours. If this were achieved, it would be the first time a planet has been directly observed in reflected light. There are also at least ten other very nearby exoplanets that could be characterised with the instrument \citep{Lovis2022} and any new understanding of these could potentially be extrapolated to more distant planets. This science case and the instrument itself act as a pathfinder for both the ArmazoNes high Dispersion Echelle Spectrograph (ANDES; \citealp{Marconi2024, Palle2025}) and the Planetary Camera and Spectrograph (PCS; \citealp{Kasper2021}) planned for the Extremely Large Telescope (ELT).

In addition to observing accreting protoplanets, as discussed in this paper, supplementary science cases include characterising solar system moons such as Io and Europa. Their on-sky sizes of 1.2 and 1.0 arcseconds are large compared to the RISTRETTO spaxel size of $\sim$37~mas, thereby enabling the resolution of wind velocities and surface features including Io's volcanoes. The spatial resolution of the instrument also promises the study of Doppler velocimetry in protoplanetary discs, thus complementing the wide array of kinematic studies already conducted using ALMA, from which the majority of our observational knowledge about discs is derived \citep{Andrews2020}.

The current specifications required to resolve Proxima~b from its host star and obtain the needed planet--star contrast are: 
\begin{itemize}
  \item Wavelength range: 620--840~nm;
  \item Spectral resolution: $R=\textrm{130,000}$--150,000;
  \item Inner working angle: 37~mas or $2\lambda/D$;
  \item Strehl ratio: $>60\%$;
  \item Stellar contrast: $1\times10^{-4}$.
\end{itemize}

Here, $R$ is the spectral resolution ($\lambda/\Delta\lambda$; \citealt{Lovis2024}). The stellar contrast is the amount of the stellar light located in the external spaxel when the star is located in the central spaxel. More detailed simulations of the wavelength dependence of this coupling can be found in \cite{Bugatti2025}. The system throughput is derived from simulations of the optical systems of the spectrograph. It contains a sum of the average transmission of the atmosphere, aluminium coating, front-end, adaptive optics and coronagraph, fibre link efficiency, and spectrograph efficiency \citep{Bugatti2024}.

\section[H alpha Sensitivity with RISTRETTO]{H$\alpha$ sensitivity with RISTRETTO}
\label{sec:Hasens}

In this section, we describe our simulations of RISTRETTO H$\alpha$ observations for theoretical protoplanets with the properties of PDS70b, PDS70c, WISPIT2b, and 2MJ1612b. We did this for both the actual separation and H$\alpha$ line fluxes of the objects, as well as for a number of theoretical protoplanets at different separations (37--600~mas). All have the same mass, radius, and spectral energy distributions. Although RISTRETTO has sensitivity between 620--840~nm, here we focus only on the H$\alpha$ peak centred on 656.28~nm.

\subsection{Input spectra}
 \label{sec:inputspec}

BT-Settl CIFIST theoretical spectra \citep{Allard2012} are available from the Spanish Virtual Observatory\footnote{\url{https://svo2.cab.inta-csic.es/theory/newov2}} at effective temperatures between 1200 and 7000~K at wavelength intervals of 0.001~\AA{} (the equivalent of $R\sim6.5\times10^6$ at H$\alpha$ $\lambda6563$). The temperature range is sufficient for these models to be used for initial estimates of both the stellar hosts and planetary companions. Based on the SED model fits with the highest Bayes factor in \citet{Wang2021}, which are consistent with the data obtained by \citet{Blakely2025}, we adopted $T_{\mathrm{eff}}=4200$~K and $\log g =4.8$ for the PDS70 host star, 1400/4.0 for PDS70b, and 1300/4.0 for PDS70c. These parameters are summarised for all targets in Table~\ref{table:planetproperties}. We then downsampled the spectra to the median resolution of RISTRETTO of $R=\textrm{140,000}$. These spectra then served as the baseline for our simulations.

On top of this underlying profile, we added the H$\alpha$ emission of both the planet and the star. In the stellar case, we used an observed PDS70 HARPS (High Accuracy Radial velocity Planet Searcher) spectrum from the ESO 3.6~m telescope \citep{Thanathibodee2020} as the stellar H$\alpha$ signal. Adding these two signals together gives an approximation of the stellar spectrum across the wavelength sensitivity (620--840~nm) of the instrument. Given the absence of appropriate real data for WISPIT2 and 2MJ1612, we added an H$\alpha$ signal of the same shape to their BT-Settl spectra. For the planetary contribution, we use the theoretical H$\alpha$ line profiles from the NLTE radiation-hydrodynamical calculations of \cite{Aoyama2018}. The line shape depends on the pre-shock hydrogen number density, $n_0$, and velocity, $v_0$ (their dependence on the full-width half-maximum can be seen in Fig.~\ref{fig:halpha_v0n0}). For brevity, we will quote $v_0$ in km/s and $\tilde{n}_0\equiv\log n_0/\textrm{cm}^{-3}$ and omit the tilde. 
We repeated the simulations for four different models: $(v_0,n_0) = (60, 12)$; $(v_0,n_0) = (60, 14)$; $(v_0,n_0) = (80, 12)$; and $(v_0,n_0) = (80, 14)$ to demonstrate the ability to distinguish between them. 

\begin{figure}[t]
\centering
\includegraphics[width=0.48\textwidth]{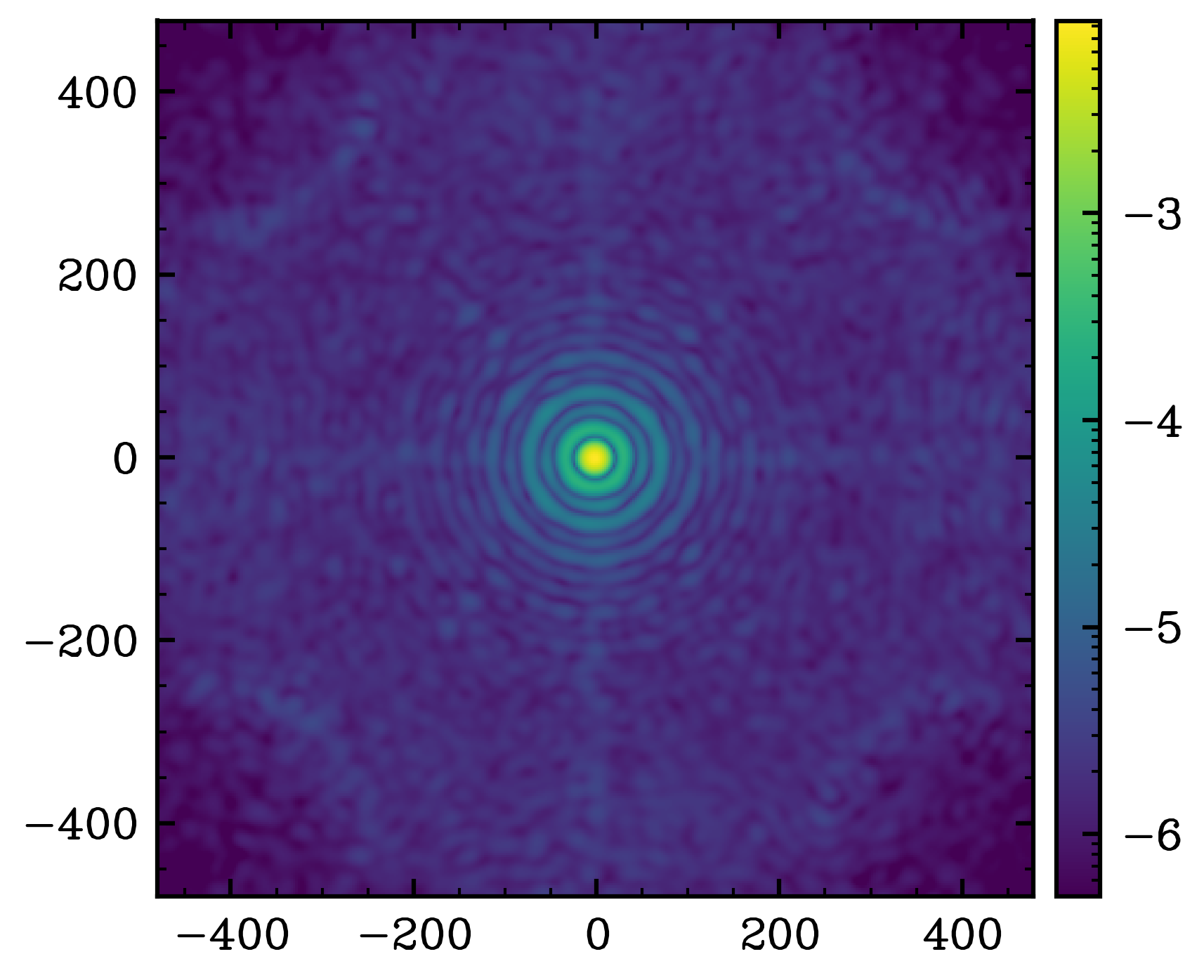}
\caption{Adaptive optics simulation of the point-spread function (PSF) with a Strehl of 0.6. At distances greater than 100~mas the background becomes comparatively uniform due to the residual speckles. We do not perform predictions between 37~mas and 100~mas due to the dominant Airy diffraction pattern.
\label{fig:airy}}
\end{figure}

\begin{figure}[t]
\centering
\includegraphics[width=0.48\textwidth]{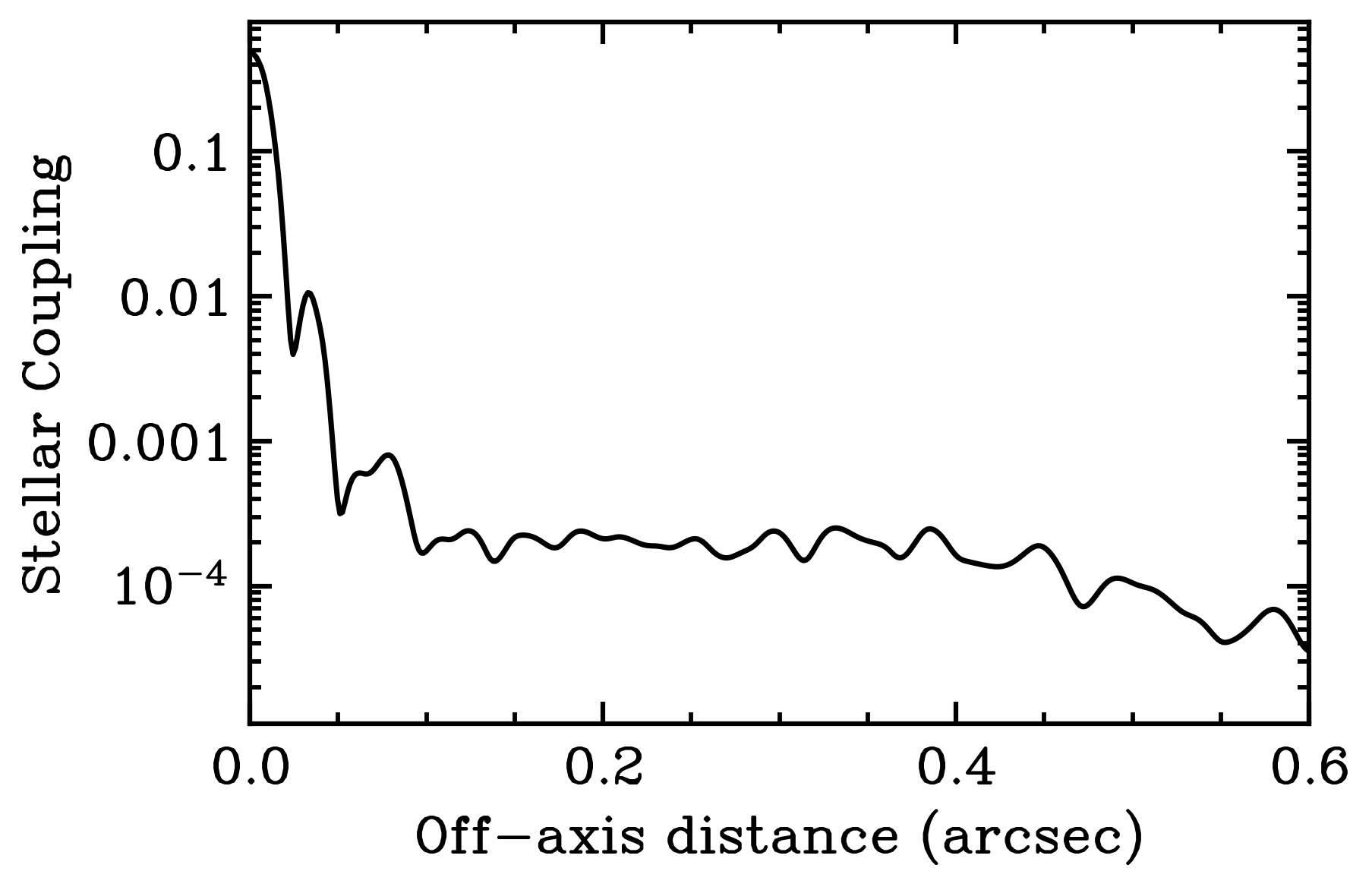}

\caption{Predicted adaptive optics coupling, $d$, for the seven spaxels when performing off-axis observations. This represents the amount of stellar light entering the central fibre when when centred on the planet as a function of separation (mas). The average stellar coupling in the centre spaxel in the range 0\farcs1--0\farcs4 is $1.9\times 10^{-4}$. When $>0\farcs4$ it is $0.9\times 10^{-4}$.
\label{fig:coupling}}
\end{figure}

These theoretical emission lines are added to the BT-Settl continuum of the planet and scaled according to the measured H$\alpha$ line flux from observations shown in Table~\ref{table:halpha_flux}.
Where there are a range of values due to differing analyses (e.g. \citealp{Haffert2019,Hashimoto2020}) or variability \citep{Close2025a}, we performed calculations for both extremes of the range, while noting the lower estimate specifically to ensure detectability. The input spectra were determined for our four targets at 12 theoretical separations (on-axis at 37~mas and off-axis at 100--600~mas in 50~mas intervals). Different observing strategies are required for on-axis and off-axis observations. This is discussed in the following section.

\begin{figure*}[th]
\centering
\includegraphics[width=0.99 
\textwidth]{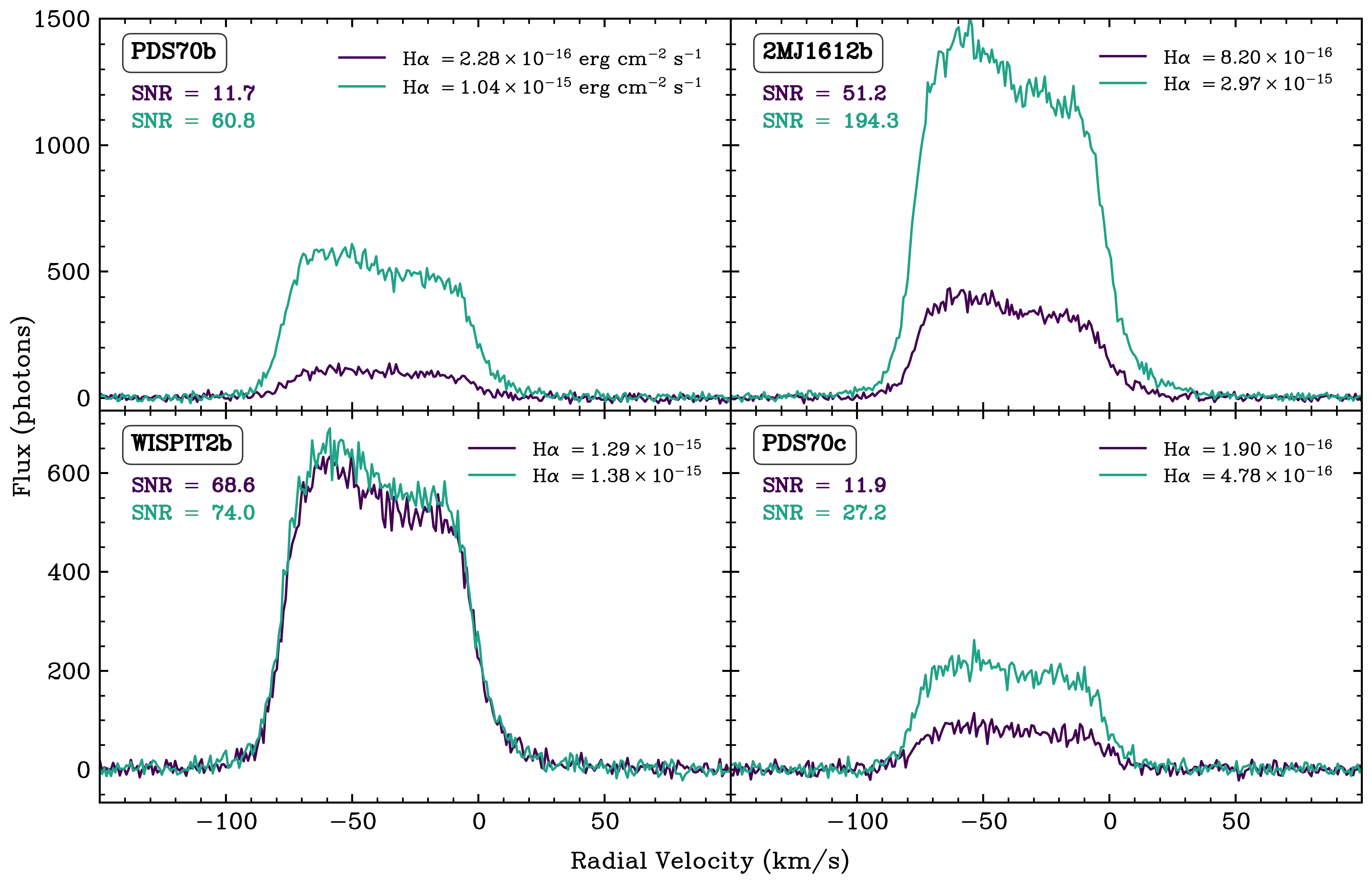}
\caption{Simulated off-axis observations of PDS70b and PDS70c, WISPIT2, and 2MJ1612b with VLT/RISTRETTO using \cite{Aoyama2018}'s model of the H$\alpha$ line (with $v_0=60\;\mathrm{km/s}$ and $n_0= 10^{14}\;\mathrm{cm^{-3}}$) with an exposure time of 3600~s. The objects are positioned at their real separation and compare the peak S/N between the upper and lower published estimates of the H$\alpha$ line flux. 
\label{fig:combinedplot_real}}
\end{figure*}

\begin{figure}[]
\centering
\includegraphics[width=0.48\textwidth]{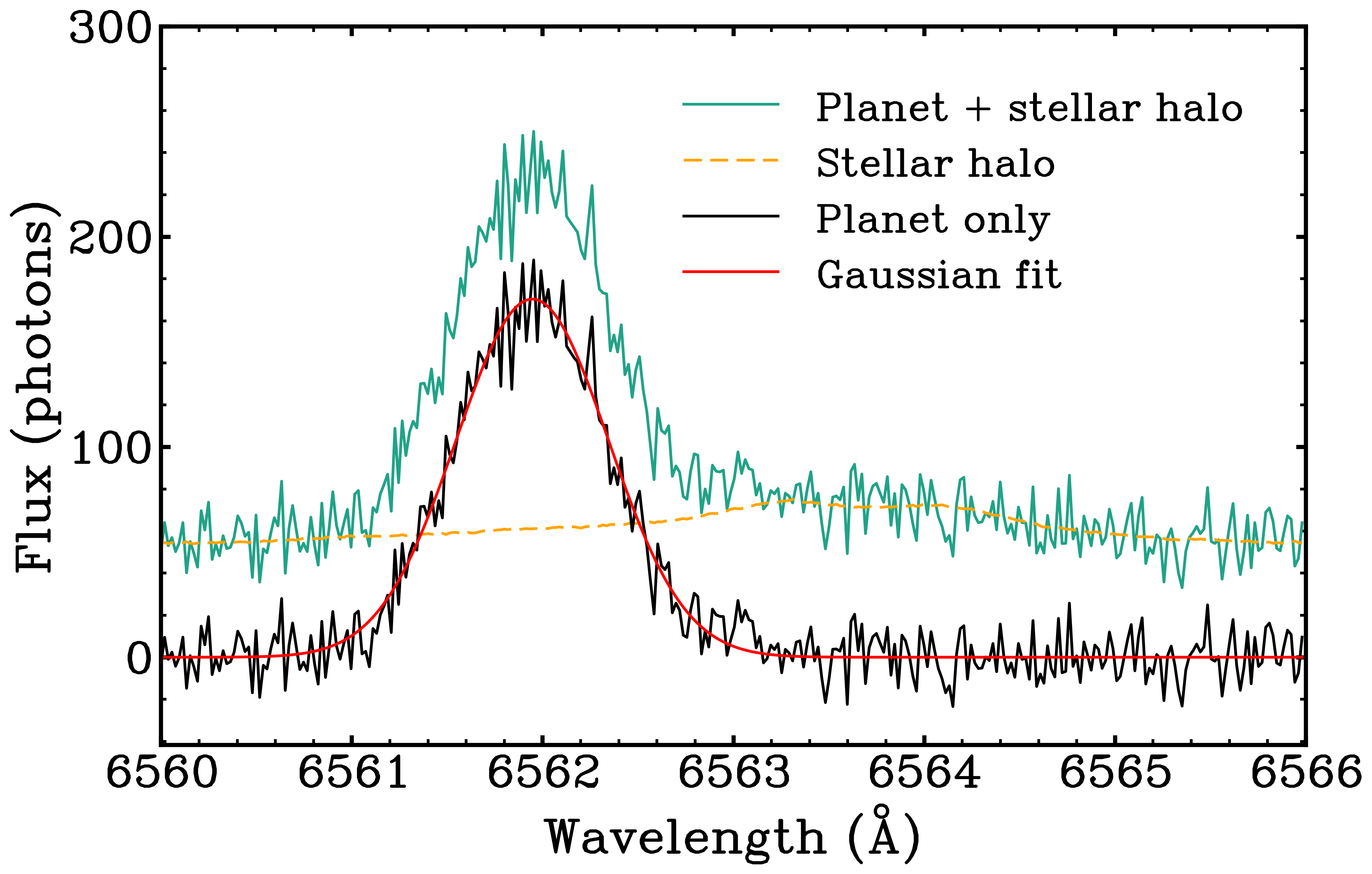}
\caption{Simulated off-axis observations of PDS70b with RISTRETTO using the \cite{Aoyama2018}
H$\alpha$ line model for $(v_0,n_0) = (60,12)$, with an exposure time of 3600 seconds. We show the photon count of the planet and stellar halo in the external spaxel (green), an estimate of the stellar halo based on an observation of the star in the centre spaxel (yellow), the resulting subtracted, planet-only signal (black), and a Gaussian fit (red).
\label{fig:pds70b_subtract}}
\end{figure}

\subsection{On-axis versus Off-axis}
\label{subsection_onaxis}
In the on-axis case where the star and planet are located within the boundaries of the central and external spaxel, respectively (see Fig.~\ref{fig:ristretto_onffaxis}), only one exposure is required. As in the case of Proxima~b, if the location of the planet in uncertain, two exposures might be needed, with a 30~degree rotation between each to ensure the planet does not sit between the spaxels \citep{Bugatti2025}. For the actual observations, it is expected that there would be some pointing error involved. As the response is not uniform across each spaxel (Fig.~\ref{fig:ristretto}), to first order, the planet coupling would drop by $\sim$50$\%$ for a centring error of 11~mas. For the purposes of our detection limit calculations, we assumed both the star and planet are located centrally in the spaxels.

For PDS70b, located at a distance of 113.4~pc, the on-axis case would be the equivalent of the planet having a separation of 3--5~AU. This corresponds to an on-sky separation of $\sim$37$\pm11$~mas. This is using the instrument as intended, with the star being located in the central spaxel and the planet in an external spaxel. For both the on-axis and off-axis case, we must correct for transmission losses through the entire system.

The overall average throughput of the instrument is estimated to be 7.7\% \citep{Bugatti2025}. This is made up of contributions from atmospheric transmission at the zenith (96.6\%), aluminium coating transmission (61.2\%), front-end transmission (68.1\%), fibre link efficiency (89.1\%), and average spectrograph efficiency (43.9\%). Combining the above losses gives an overall efficiency of 15.7\% $\times\;d$, where d is the coupling through the AO and coronagraph. This coupling is dependent on target elevation and the alignment of the target star and planet within the spaxel array. It describes the amount of light that effectively enter the fibres. Considering a median seeing of 0\farcs76 (at the zenith), median elevation of $56\degr$ and a Strehl ratio of $\sim$60\%, we estimate the planet coupling in the six external fibres of the IFU to be $d_\mathrm{P}=0.45$, and the stellar coupling in the same fibres to be $d_\mathrm{\star}=3.5\times10^{-4}$. This stellar coupling quantifies the stellar halo in the external spaxels when the star is in the central spaxel. The coupling in this central fibre is also estimated to be $0.45$, although there a small coupling boost is predicted as $0.51$ when using the PIAA/coronagraph.

The off-axis case is required for simulations of the real objects (all with separations of 142--310~mas) and our theoretical cases. We did not examine the region between 37~and 100~mas due to the Airy ring diffraction pattern (see Fig.~\ref{fig:airy}), although science may eventually be possible in the region once the instrument is completed and the non-linearities are properly characterised. The contrast will also be at its worst in this region due the proximity of the host star and inability to use the coronagraph.

For the off-axis observations, we needed to adopt a dual exposure strategy, centring both the star and the planet on the central spaxel (as per Fig.~\ref{fig:ristretto_onffaxis}). To maintain consistency when performing stellar halo subtractions on the planetary exposures, both objects should be observed using the same spaxel. Two exposures would be required: a shorter minutes-long exposure for the host star and a longer exposure (up to 1~hour, which is the readout limit of the detector) for the protoplanet. Figure~\ref{fig:coupling} shows the predicted brightness of the stellar halo as a function of off-axis distance. The signal when the planet is located in the centre spaxel will therefore be the sum of the planetary signal, scaled by $0.157 \times0.45$, and the stellar halo, scaled by $0.157\times d$, where $d$ is the coupling shown in Fig.~\ref{fig:coupling}.

\subsection{PyEchelle Simulations}

For the input spectra described in Sect.~\ref{sec:inputspec}, we converted the flux at the surface of the objects (in erg\; s$^{-1}$\; cm$^{-2}$~\AA$^{-1}$) to photons~s$^{-1}$ at the surface of the telescope. We used the \texttt{python}-based simulation tool PyEchelle \citep{Sturmer2018} to generate realistic 2D echelle spectra. PyEchelle accounts for geometric transformations on the detector plane due to scaling, rotation, shearing and translation and uses transformation matrices to apply these factors to a wavelength-dependent input. It takes into account readout noise, photon noise, bias, and radial velocity perturbations. It also uses wavelength-dependent point spread functions (PSFs) to properly account for optical aberrations. The throughput correction described in the previous section is not included within the pipeline and is applied manually to the input before processing. As in \cite{Bugatti2025}, we set the read-out noise to 3~electrons in line with the CCD specifications, bias to 250 and ignore the minimal dark current that is estimated to be 0.5~electrons/pixel/hour. Poisson photon noise is added automatically by PyEchelle.

\begin{figure*}[t!]
\centering
\includegraphics[width=0.99 
\textwidth]{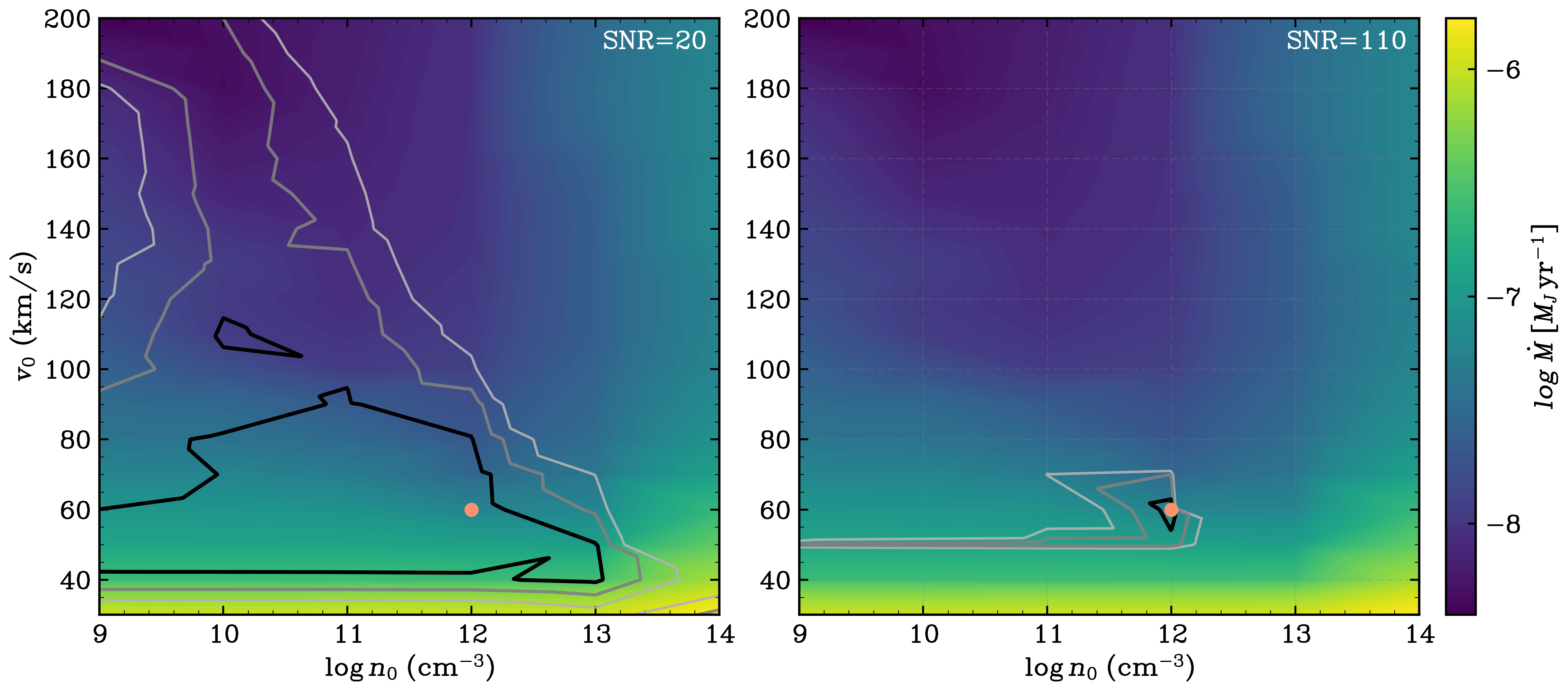}
\caption{Predictions of the mass accretion rate, $\dot{M}$, based on the $(v_0,n_0)=(60,12)$ planetary H$\alpha$ model of \cite{Aoyama2018} calculated using Eq.~\ref{eq:accretionrate}. The greyscale contours show 1--3$\sigma$ constraints. This figure indicates the improvement in constraints due to the increase in S/N from 20 (left) to 110 (right). These S/N values are based on our PyEchelle simulations of RISTRETTO observations of PDS70b.
\label{fig:mdot}}
\end{figure*}

\begin{figure}[t]
\centering
\includegraphics[width=0.48 
\textwidth]{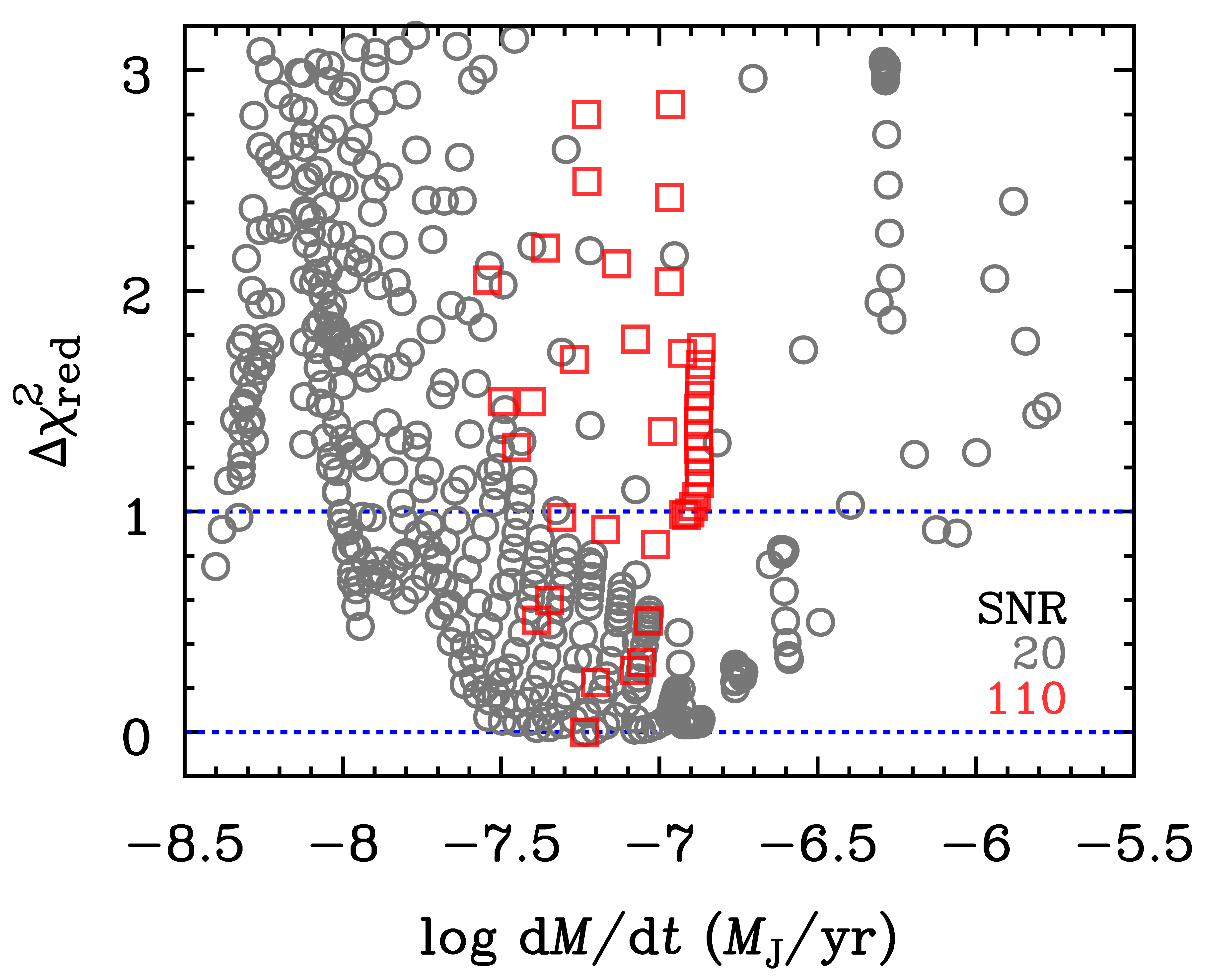}
\caption{Constraints on the accretion rate, $dM/dt$, for simulated observations with S/N of 20 (grey) or 110 (red). Best-fit values and error bars are determined from the shape of the distribution where $\Delta\chisqred\leqslant1$.   
\label{fig:chi2red}}
\end{figure}

\begin{figure*}[h]
\centering
\includegraphics[width=0.99 
\textwidth]{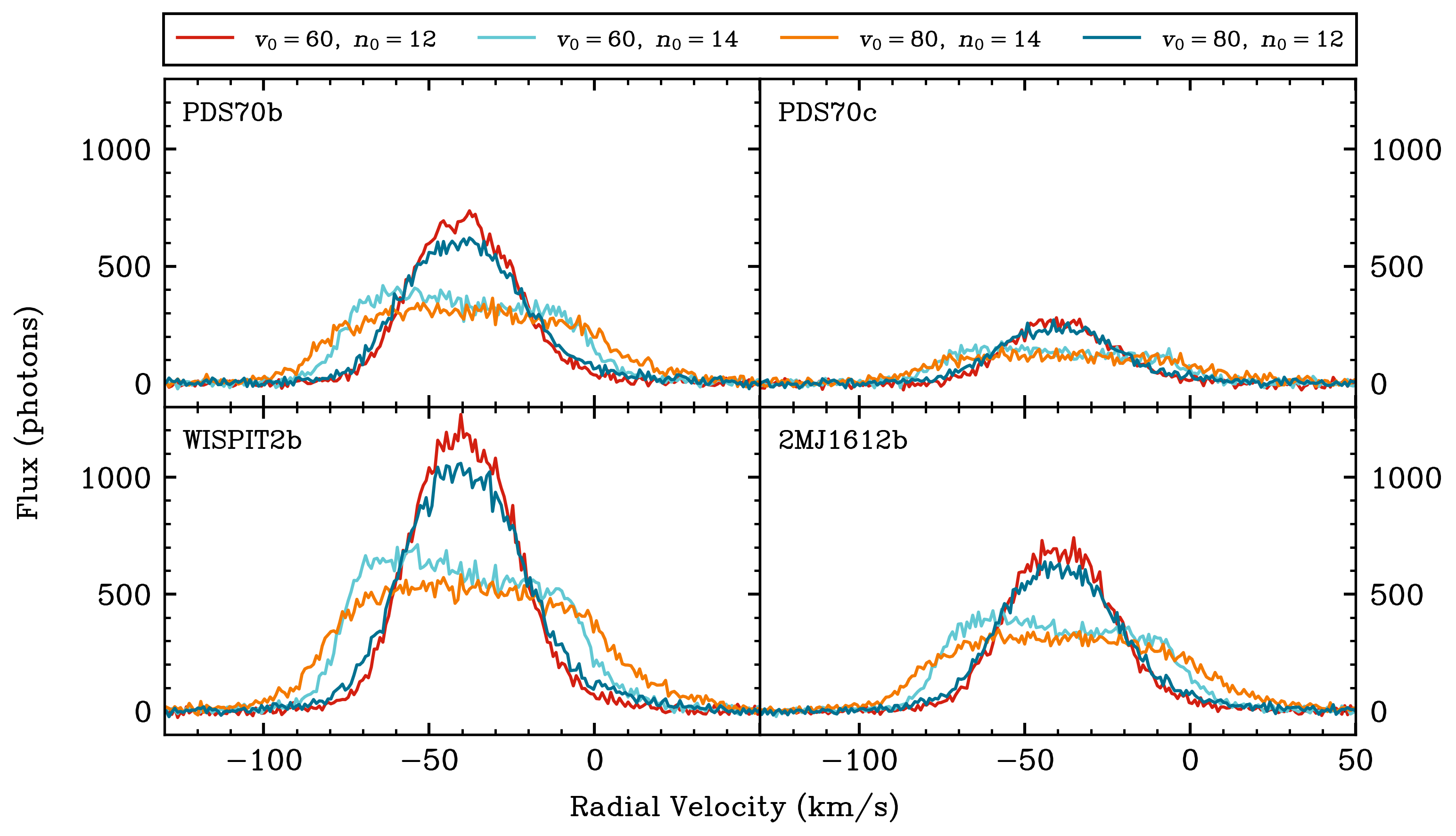}
\caption{Simulated H$\alpha$ observations of PDS70b, PDS70c, WISPIT2b, and 2MJ1612b from 1 hour exposures with VLT/RISTRETTO. The coloured curves correspond to four different H$\alpha$ models \citep{Aoyama2018} with varying pre-shock velocity, $v_0$, and hydrogen number density, $n_0$. The visible difference in line shape shows how these underlying accretion parameters can be determined from the FWHM line width (see Fig.~\ref{fig:halpha_v0n0} for a visual representation of this relationship). Since the emission line is fully resolved the FWHM can be determined up to a 0.3\% precision.}
\label{fig:modelcompare}
\end{figure*}

\begin{figure*}[t]
\sidecaption
\includegraphics[width=0.58\textwidth]{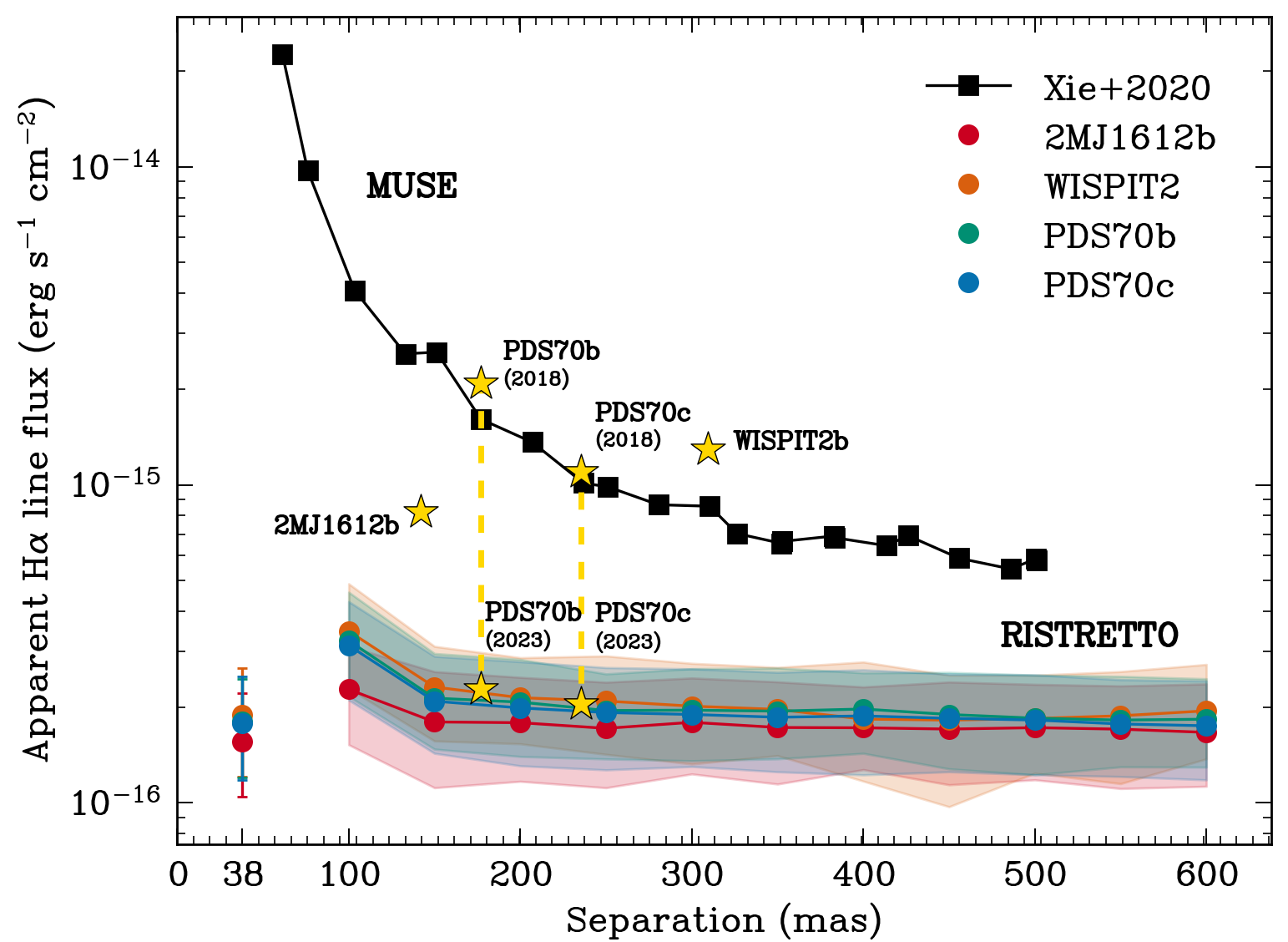}
\caption{5$\sigma$ H$\alpha$ detection limits for theoretical protoplanets with the properties of PDS70b, PDS70c, WISPIT2, and 2MJ1612b at separations between 37-600~mas when observed with RISTRETTO. The shaded area and error bar at 37~mas indicate the range of limits due to the different models ($v_0=60,80$; $n_0=12,14$). In black: $5\sigma$ limits for PDS70b observed with MUSE (30~min exposure time) from \cite{Xie2020}. Yellow stars show the locations of the real objects, with their flux estimates given in Table. \ref{table:halpha_flux}. The upper estimates for PDS70b and PDS70c are from \cite{Xie2020}, while the lower estimates are from more recent 2023 MagAO-X observations described in \cite{Close2025a} demonstrating the intrinsic variability in both objects.
\label{fig:ristretto-limits}}
\end{figure*}

\subsection{Results}

The simulated off-axis observations of our four targets at their actual separations, centred on H$\alpha$ at 6562.82~\AA, with input planetary model parameters $(v_0, n_0) = (60,14)$, are shown in Fig.~\ref{fig:combinedplot_real}. This figure shows the pure planetary spectrum in the external spaxel following the subtraction of the residual stellar signal (scaled by the predicted coupling). For comparison, Fig.~\ref{fig:pds70b_subtract} shows the method of subtraction, but for the model $(v_0, n_0) = (60,12)$, which has a narrower line shape more similar to a Gaussian. In all cases, we utilised a theoretical H$\alpha$ spectrum of the planet and therefore the planetary line width may differ from what is observed. We used both the maximum and minimum H$\alpha$ line fluxes predicted in the literature to show the range of plausible signal-to-noise ratios (S/Ns). The S/N based on the lowest flux estimates in Table~\ref{table:halpha_flux} is sufficient for detection and it would also be valid for broader line widths \citep{Aoyama2019}.

Assuming the maximum estimated H$\alpha$ line flux, $F=(10.4\pm1.6)\times10^{-16} \mathrm{\;erg\; s^{-1}\; cm^{-2}}$, we predicted RISTRETTO observations of PDS70b using the $(v_0, n_0) = (60,14)$ model would have a peak SN/ of 60.8 after a 1 hour exposure. For the lower line flux, $F=(2.28\pm0.26)\times10^{-16} \mathrm{\;erg\; s^{-1}\; cm^{-2}}$, the S/N is 11.7. A correction due to the radial velocity of the H$\alpha$ emission was applied (4.3~km/s; \citealt{Haffert2019}), which allows us to offset the planetary H$\alpha$ from the stellar and makes  it easier to isolate the planetary signal. The resulting  H$\alpha$ line has a FWHM of $1.47\pm 0.03$~\AA{} ($66.8 \pm 1.5$~km/s). The thinner line shape shown in Fig.~\ref{fig:pds70b_subtract}, with $n_0=12$, results in a higher S/N of 20.3--110.3. For this $(60,12)$ case, we estimated $v_0$ and $n_0$ using the relationship shown in Fig.~\ref{fig:halpha_v0n0} and we subsequently placed constraints on the mass accretion rate, $\dot{M}$, as per Eq.~\ref{eq:accretionrate}. The distributions of these parameters are shown in Fig.~\ref{fig:mdot} and indicate that there is a significant dependence on S/N in breaking the $v_0$--$n_0$ degeneracy. By minimising $\chisqred=\chisqred(n_0,v_0)$, we estimated log$\dot{M}=-7.2\pm0.3$ for S/N=110 and log$\dot{M}=-7.3^{+1.3}_{-0.7}$ for S/N=20. These numbers are derived from the distributions between $\Delta\chisqred=0-1$ shown in Fig. \ref{fig:chi2red} and show that our achievable S/N is sufficient to place constraints on the mass accretion rate.

RISTRETTO observations of the fainter PDS70c are predicted to yield a $\textrm{S/N}=12$ for the lowest line flux of $F=(1.9\pm0.32)\times10^{-16} \mathrm{\;erg\; s^{-1}\; cm^{-2}}$ with a corresponding H$\alpha$ $\textrm{FWHM}=1.45\pm 0.04$~\AA{} ($66.3 \pm 1.8$~km/s). The upper line flux results in $\textrm{S/N}=27.2$. For the same object, but with the $(60,12)$ model the S/N varies between 20.5 and 47.8. Even in the case of the broadest width test case of $(80,14)$, the object is still observable with a $\textrm{S/N}=9.7$. 

Considering only the lower H$\alpha$ line flux estimate from \cite{Li2025a} for 2MJ1612b yields $\textrm{S/N}=51.2$ and the $\textrm{FWHM}=1.45\pm 0.03$~\AA{} ($66.4 \pm 1.4$~km/s). This flux of $8.2\times10^{-16} \mathrm{\;erg\; s^{-1}\; cm^{-2}}$ was measured to be over three times greater just three days prior ($29.7\times10^{-16} \mathrm{\;erg\; s^{-1}\; cm^{-2}}$). This sees an increase to S/N=194.3 where FWHM=$1.47\pm 0.03$~\AA{} ($67.0 \pm 1.4$~km/s). The (60,12) case results in a S/N range of 89.8-338.4. In the final case of WISPIT2b, S/N=68.6-73.9 with FWHM= $1.48\pm 0.03$~\AA{} ($67.5 \pm 1.4$~km/s) for the (60,14) model. For (60,12), $\textrm{S/N}=130.2$--131.1. The variance is small here thanks to the narrow range of flux estimates, which only differ by $0.9\times10^{-16} \mathrm{\;erg\; s^{-1}\; cm^{-2}}$.  For all of the objects considered here, the same theoretical model of the planetary H$\alpha$ line with is being used and they hence have near identical widths. The shape (and, hence, the FWHM) is particularly dependent on $n_0$, with a larger $n_0$ resulting in a broader line width. Figure~\ref{fig:modelcompare} compares shape of the four models.

We performed PyEchelle simulations for all four objects at 12 separations at 18 measured and theoretical H$\alpha$ line fluxes for a total of 864 simulated spectra. The end result of this procedure is the VLT/RISTRETTO 5$\sigma$ detection limit in terms of H$\alpha$ line flux as a function of separation shown in Fig.~\ref{fig:ristretto-limits}. The average lower limit for all four objects between 150--600~mas is $1.4\times10^{-16} \mathrm{\;erg\; s^{-1}\; cm^{-2}}$. The average detection limit at 100~mas is $3\times10^{-16} \mathrm{\;erg\; s^{-1}\; cm^{-2}}$. The on-axis case at 37~mas has the best sensitivity at $1.8\times10^{-16} \mathrm{\;erg\; s^{-1}\; cm^{-2}}$. This is because this geometry takes full advantage of the seven-spaxel design of RISTRETTO. This is also the area of parameter space inaccessible to instruments such as MUSE due to the lack of spatial resolution and the unfavourable stellar contrast ratios.

\section{Conclusion and discussion}
 \label{sec:concdisc}
 
In this paper, we demonstrate the capability of RISTRETTO to detect currently accreting protoplanets embedded in gaseous discs. We show that the H$\alpha$ emission from protoplanets PDS70b and PDS70c, WISPIT2, and 2MJ1612b are detectable with the maximum 1 hour exposure time, in the worst-case scenario of the lowest published line flux, with a S/N of 11.9, 11.7, 68.6, and 51.2. respectively. We also show that theoretical protoplanets with their properties at separations of 37 and 100--600~mas, corresponding to 3--5 and 10--70~AU at the distance of PDS70b (113.4~pc), are observable. This distance is near the lower limit of the closest star-forming regions such as the Ophiuchus (140~pc) and Taurus (140~pc) molecular clouds. The region 37--100~mas may become accessible with rigorous modelling of the adaptive optics PSF once the instrument is complete.

While the four planets considered here $-$ and any other embedded protoplanets that may be discovered prior to RISTRETTO's expected installation on the VLT (currently $\sim$2030) $-$ are priority targets, it seems feasible to conduct a search for protoplanets with $\sim$37~mas (3--5~AU) separations (i.e. the on-axis case where the planet and star are observed simultaneously). The giant planet occurrence rate peaks in these orbits, with \cite{Fulton2021} reporting an occurrence rate of 14.1 giant planets per 100 stars in the region 2--8~AU compared with 8.9 planets beyond this. There is the caveat that a blind survey of planets in these regions would be restricted by the system geometry. Core accretion theory \citep{Pollack1996, Emsenhuber2025} also predicts this region at and beyond the snow-line is where most giant planets form. Such a study would be complementary to microlensing studies of planets around M~dwarfs, where theory may underestimate observations by a factor of ten \citep{Suzuki2018}. However, a rigorous treatment of the scientific yield of this blind survey is beyond the scope of this study.

The low-resolution H$\alpha$ observations of PDS70b with MUSE shown in \cite{Aoyama2019} are satisfyingly fit using a Gaussian, although there is some broadening in the wings. Figures~\ref{fig:pds70b_subtract} and~\ref{fig:modelcompare} shows that this approximation is adequate for emission with smaller hydrogen number density ($n_0$), but breaks down when fitting the wider, non-uniform profiles (e.g. $n_0 > 14$), particularly due to the requirement of obtaining varying line widths (e.g. 10\%, 90\%) to break the line width and $n_0$-$v_0$ degeneracy. More complex line shape models are therefore required when observing at high resolution. The potential variability in H$\alpha$  line flux also suggests that time series observations might be necessary to obtain a full picture of the underlying accretion processes.

In summary, we show that the RISTRETTO instrument will make it possible to obtain very high spatial and spectral resolution observations of accreting protoplanets. It will enable us to see the spectral fingerprints of different physical processes imprinted on the line profile. This will yield crucial novel constraints for planet formation theory, such as the planet gas accretion rate, the spatial origin of the shock emission (i.e. protoplanet surface, circumplanetary disc), different accretion geometries (spherical, polar, magnetospheric), planetary magnetic fields strength, and $-$ via molecular lines in incoming gas absorbing part of line $-$ the composition.

\begin{acknowledgements}
Parts of this work have been carried out within the framework of the NCCR PlanetS supported by the Swiss National Science Foundation under grants 51NF40\_182901 and 51NF40\_205606.
G.-D.M.\ acknowledges the support of the Deutsche Forschungsgemeinschaft (DFG) through grant MA~9185/2-1.
C.M.\ acknowledges the support from the Swiss National Science Foundation under grant 200021\_204847 ``PlanetsInTime''.

This publication made use of the ``Theoretical Spectra'' service of VOSA (\url{https://svo.cab.inta-csic.es}; \citealp{Bayo2008}), developed under the Spanish Virtual Observatory (SVO) project funded by MCIN/AEI/10.13039/501100011033/ through grant PID2020-112949GB-I00. VOSA has been partially updated by using funding from the European Union's Horizon 2020 Research and Innovation Programme ``EXOPLANETS-A'' (Nr.~776403). The text in this paper was written by humans.
\end{acknowledgements}

\bibliographystyle{aa_url}
\bibliography{library}

\end{document}